\documentclass{PoS}

\title{Four fermion condensates in $SU(2)$  Yang-Mills-Higgs theory on a lattice}

\ShortTitle{Symmetric mass generation}

\author{\speaker{Nouman Butt} and Simon Catterall \thanks{smcatterall@gmail.com}\\
Department of Physics, Syracuse University, Syracuse, New York 13244, United States\\
        E-mail: \email{ntbutt@syr.edu}}

\usepackage[T1]{fontenc} 
\usepackage{amsmath}
\usepackage{breqn}
\usepackage{amssymb}
\usepackage{bbold}
\newcommand{\beq}{\begin{equation}}
\newcommand{\eeq}{\end{equation}}
\usepackage{comment}

\abstract{We study a model of four reduced staggered fields transforming in the bifundamental representation of a $SU(2)\times SU(2)$ symmetry group where just one of the SU(2) factors is gauged. This field content and symmetries are similar to a Higgs-Yukawa model that has been studied recently. The key observation in the latter work is that fermions acquire masses at strong coupling via the formation of a symmetric four fermion condensate in contrast to the more usual symmetry breaking bilinear condensate seen in eg. NJL models. The current work attempts to see whether this structure survives when the four fermi interactions are replaced by gauge interactions and to explore the resulting phase diagram}

\FullConference{The 36th Annual International Symposium on Lattice Field Theory - LATTICE2018\\
		22-28 July, 2018\\
		Michigan State University, East Lansing, Michigan, USA.}

\begin{document} 
\maketitle
\flushbottom
\section{Introduction}

The motivation to study this particular lattice model stems from previous studies~\cite{Catterall:2016dzf,Ayyar:2016lxq,Catterall:2015zua,Ayyar:2014eua,Ayyar:2015lrd} of an $ SU(4) $ invariant four-fermion model in three dimensions which possesses an unusual phase structure in which fermions acquire
masses at strong coupling {\it without} forming a symmetry breaking bilinear condensate. A single continuous phase transition with non mean field
exponents separates this strong coupling phase from a massless free fermion
phase.
In four dimensions it appears that
a narrow broken phase separates the massless and massive phases \cite{Schaich:2017czc} but 
recent work \cite{Butt:2018nkn} provided evidence that
this broken phase can be evaded in a Higgs-Yukawa generalization of the model in which the auxiliary field used to generate the four fermion interaction is given a kinetic
term.
In this expanded phase diagram there is evidence that once again a direct transition between massless and massive phase is possible. Since there is no symmetry breaking this novel transition eludes a Landau-Ginzburg description in terms of a local order parameter. Instead the transition
is argued to result from  a proliferation of topological defects in the scalar field \cite{Catterall:2017ogi}.

The phenomenon of symmetric mass generation has also received a great deal of
interest in condensed matter physics ~\cite{Slagle:2014vma,Morimoto:2015lua,He:2016sbs} and in three dimensions this phenomenon was conjectured to be described by a gauge theory ~\cite{You:2017ltx}.

The Higgs-Yukawa models that have been used to generate this novel structure are invariant under an $SO(4)$ symmetry and utilize  a scalar field which transforms in the adjoint representation of one of the $ SU(2)$ factors
making up $SO(4)=SU(2)\times SU(2)$. In the rest of this paper we call this $SU_{+}(2)$. This scalar is a singlet under the other factor which we call $SU_{-}(2)$.  
It is then rather natural to imagine replacing the adjoint scalar field with a corresponding gauge field and ask whether the resulting gauge theory is capable of generating a four fermion condensate even in the absence of a Yukawa coupling. In practice we have considered a model containing both scalars and gauge fields and tried to map out the resulting phase diagram.

We expect for small Yukawa coupling that the theory is in a confined phase while
for weak gauge coupling and large Yukawa coupling one might expect to see the four fermion condensate which can be interpreted as a Higgs phase of the theory. 
it is plausible that one might expect to see a line of phase transitions separating these two phases the two dimensional phase diagram spanned by the Yukawa and gauge couplings.

\section{Fermion kinetic term}

Consider staggered fermions in the bifundamental representation of an $ SU(2) \times SU(2) $ symmetry. The fermions transform under a general gauge transformation as
\beq
 \psi \to G \psi H^{\dagger} 
  \eeq
where $ G \in SU_{+}(2) $ and $ H \in SU_{-}(2) $.  The above transformation with left and right action of the group is equivalent to the standard tensor transformation 
\beq
\psi^{Aa} = G^{AB} H^{*ab} \psi^{Bb}
\eeq
To construct the model we start with the full staggered action given in ~\eqref{eq:stag} gauged under both $SU(2)$ factors:
\begin{equation}
\label{eq:stag}
S_F = \sum_{x,\mu} \frac{1}{2} \eta_{\mu}(x)Tr [ \psi^{\dagger}(x) U_{\mu}(x) \psi(x+\mu)V_{\mu}^{\dagger}(x) - \psi^{\dagger}(x) U^{\dagger}_{\mu}(x-\mu)\psi(x-\mu) V_{\mu}(x-\mu)] 
 \end{equation}
This action is invariant under the following gauge transformations
\beq
\label{eq:x}
\begin{split}
\psi(x) \to G(x) \psi(x) H^{\dagger}(x) \\
U_{\mu}(x) \to G(x)U_{\mu}(x) G^{\dagger}(x+\mu) \\ 
V_{\mu}(x) \to H(x+\mu) V_{\mu}(x) H^{\dagger}(x) 
\end{split}
\eeq
The only single site gauge invariant mass term for fields in the bifundamental representation is $ Tr (\psi^{\dagger}\psi) $ which vanishes on account of the Grassmann nature of the fields. We can, however, construct a gauge invariant four-fermion term given in~\eqref{eq:four}
\beq
\label{eq:four}
Tr(\psi^{\dagger}\psi \psi^{\dagger}\psi) 
\eeq
which is non-zero and yields the usual four fermion term.

\section{Imposing the reality condition}

The equivalence between this model and the original four fermi models requires the imposition of two further constraints. First we need to impose a reality condition on the fermions to reduce to four
real degrees of freedom and second we will eventually set the gauge coupling for $SU_{-}(2)$ to zero and set $V_\mu(x)=I$. For the moment
tet us focus the first of these which is equivalent to imposing the
constraint
 \beq
  \psi^{\dagger} = \sigma_2 \psi^{T} \sigma_2
 \eeq
 This implies that the fermion field can be written in terms of four real components $\chi_\mu,\mu=1\ldots 4$
 \beq
 \psi=\sum_\mu \chi_\mu \sigma_\mu
 \eeq
 where  $ \sigma_{\mu} = (I,i\sigma_{i}) $ and the original $SO(4)$ fields can be recovered using the relation
 \beq
 \chi_\mu=\frac{1}{2} Tr \left(\sigma_\mu\psi\right)
 \eeq
Notice that the previous four fermion term $Tr\left(\psi^\dagger \psi \psi^\dagger \psi \right)$ reduces to the simple form $\chi_1\chi_2\chi_3\chi_4$ after this.
Making use of this condition we can write the action as 
\beq
\label{eq:pseudo-real}
S_F = \sum_{x,\mu} \frac{1}{2} \eta_{\mu}(x)Tr [ \psi^{T}(x) \mathcal{U}_{\mu}(x) \psi(x+\mu)\mathcal{V}_{\mu}^{T}(x) - \psi^{T}(x) \mathcal{U}^{T}_{\mu}(x-\mu)\psi(x-\mu) \mathcal{V}_{\mu}(x-\mu)] 
\eeq
where 
\beq
\label{eq:newlinks}
\begin{split}
\mathcal{U}_{\mu}(x) = \sigma_2 U_{\mu}(x) \\
\mathcal{V}_{\mu}(x) = -V_{\mu}(x) \sigma_2
 \end{split}
\eeq
This fermion operator  is manifestly anti-symmetric 
\beq
\label{eq:operator}
M = \sum_{\mu} \mathcal{U}_{\mu}(x) \delta(x+\mu,x) \mathcal{V}^{T}_{\mu}(x) - \mathcal{U}^{T}_{\mu}(x-\mu) \delta(x-\mu,x) \mathcal{V}_{\mu}(x-\mu) 
\eeq
where $ \mathcal{V} $ acts from the right. This form of the action  leads to a Pfaffian rather than a determinant after the fermion integration. Moreover this fermion operator inherits the reality condition \beq
\label{eq:real-op}
 M^{*} = \sigma_2 M \sigma _2
 \eeq
Combining anti-symmetry and pseudo-reality we expect $M$ to exhibit a quartet of complex eigenvalues $ \left(\lambda,\bar{\lambda}, -\lambda ,- \bar{\lambda}\right) $. This guarantees that the fermion operator will
have generically possess a real, positive definite Pfaffian. The exception
to this will be if the operator develops a purely real eigenvalue. We have
not observed this to be the case in our work. 

In fact the situation is even better than this. Let us now  return to the second constraint on the model. The model we are finally
interested has only one set of gauge fields corresponding to $SU_{+}(2)$. If
the gauge links corresponding to $SU_{-}(2)$ ( $V_{\mu}(x) = \mathbb{1}$ ) are set to unity all the eigenvalues of $M$ are doubled and positivity is then completely guaranteed.

\section{Adding Yukawa interactions}
To facilitate the formation of a gauge invariant four fermion interaction  we add the term given in~\eqref{eq:four} after which the action is
\begin{dmath}
S_F = \frac{1}{2}\sum_{x,\mu}\eta_{\mu}(x)Tr [ \psi^T(x) \mathcal{U}_{\mu}(x) \psi(x+\mu)\sigma_2 ] - \frac{1}{2}\sum_{x,\mu} \eta_{\mu}(x)Tr [\psi^T(x) \mathcal{U}^T_{\mu}(x-\mu) \psi(x-\mu)\sigma_2 ]  \\ + \frac{G^2}{4} \sum_{x}Tr (\psi^{T}\sigma_2\psi\sigma_2\psi^{T}\sigma_2\psi\sigma_2) 
\end{dmath}
As usual an action quadratic in fermionic variables can be achieved if we introduce an auxiliary field. In this case there are two fermion bilinears defined in~\eqref{eq:bilin} each of which transforms in the adjoint representation under one of the $ SU(2) $'s and is a singlet under the other. Here we have used $ \psi^{\dagger} $ rather than $ \psi^{T} $ to exhibit more clearly the transformation properties of each bilinear.
\beq
\begin{split}
\psi^{\dagger}\psi \to H \psi^{\dagger}\psi H^{\dagger} \\
\psi\psi^{\dagger} \to G \psi\psi^{\dagger} G^{\dagger}
\end{split}
\label{eq:bilin}
\eeq
The two possible auxiliary fields $ \phi(x) $ and $ \sigma(x) $ 
must transform as
\beq
\begin{split}
 H \sigma H^{\dagger} \\
 G \phi G^{\dagger}
\end{split}
\eeq
Since in the end we will choose to gauge only $SU_{+}(2)$ we
choose to include only the $\phi$ auxiliary field in our work.
\begin{dmath}
\label{eq:aux}
S_F = \frac{1}{2}\sum_{x,\mu} \eta_{\mu}(x)Tr [ \psi^T(x) \mathcal{U}_{\mu}(x) \psi(x+\mu) \sigma_2] - \frac{1}{2}\sum_{x,\mu} \eta_{\mu}(x)Tr [\psi^T(x) \mathcal{U}^T_{\mu}(x-\mu) \psi(x-\mu)\sigma_2] + \frac{G}{2} \sum_{x} Tr [\psi^{T}(x)\sigma_2 \phi(x)\psi(x)\sigma_2]  + \frac{1}{2}\sum_{x} Tr [ \phi^2 ] 
\end{dmath}
Notice that the field $\phi$ is strictly forbidden from picking up a vev as that would imply spontaneous breaking of gauge symmetry which is forbidden
in a lattice gauge theory\footnote{Unfortunately while $\phi$ is a singlet
Under $SU_{-}(2)$ it does not obey the reality condition so that octct structure of eigenvalues of the fermion operator is reduced once again to
quartets.}.
To test for spontaneous breaking of the global $SU_{-}(2)$ symmetry
we can also add a gauge invariant mass term given in~\eqref{eq:mass} to the action 
\beq
\label{eq:mass}
m \sum_{x}Tr[ \sigma_3 \psi^{\dagger}\psi ]
\eeq
This term explicitly breaks $SU_{-}(2) \to U(1) $. Finally, for the
gauge part of action we have employed the standard Wilson action 
\begin{equation}
S_{G} = \sum_{x}\sum_{\mu < \nu} -\frac{\beta}{2N} Tr [ U_{\mu\nu }(x) + U^{\dagger}_{\mu\nu}(x) ]
\end{equation}

\section{Reduction to $ SO(4) $ model}
To connect this model to the original $ SO(4) $ model studied in ~\cite{Catterall:2016dzf} we consider the limit $ \beta \to \infty $ which
allows us to set $ U_{\mu}(x) = \mathbb{1} $. The fermion operator then reduces to a symmetric difference operator and the action becomes
\beq
S= \sum_{x,\mu} \frac{1}{2}Tr [ \psi^{T}\sigma_2(\eta.\Delta)\psi\sigma_2 ] + \frac{G}{2} \sum_{x} Tr [\psi^{T}(x)\sigma_2 \phi(x)\psi(x)\sigma_2] +\frac{1}{2} \sum_{x} Tr [\phi^2(x) ] 
\eeq
Transforming in the fundamental of $ SO(4) $ one finds
\begin{dmath}
S = \sum_{x,\mu} \frac{1}{2} \chi^a (\eta.\Delta) \delta^{ab} \chi^{b} + \frac{G}{2}\sum_{x} [ \phi_1 ( \chi_1\chi_2 + \chi_3\chi_4) + \phi_2 (\chi_1\chi_3 + \chi_2\chi_4) + \phi_3(\chi_1 \chi_4 + \chi_2 \chi_3) ]+ \frac{1}{2} \sum_{x} (\phi_i)^2
\end{dmath}
Notice that the fermion bilinears
appear in the self-dual representation of $ SO(4) $ 
in these variables. This projection on
self-dual fields can then be transferred 
to the auxiliary field and after performing the
trace over $SU(2)$ indices we recover the $SO(4)$ invariant action studied in~\cite{Catterall:2016dzf}
\begin{equation}
\label{eq:z}
S = \sum_{x,\mu} \frac{1}{2} \chi^a [ (\eta.\Delta) \delta^{ab} + \frac{G}{2}\phi^{ab}_{+} ] \chi^b  + \frac{1}{2} \sum_{x} (\phi^{ab}_{+})^2
\end{equation}

\section{Numerical Results}

Now we come to some preliminary results obtained by using the RHMC algorithm to simulate the model in~\eqref{eq:aux}. Our code utilizes the MILC libraries
to allow for efficient parallelization to allow for studies on
large lattice. However, our results so far have been confined 
to a small volume $ 4^4 $ lattice suitable for testing and validation
of the code. 

\begin{figure*}
\includegraphics[width=.5\textwidth]{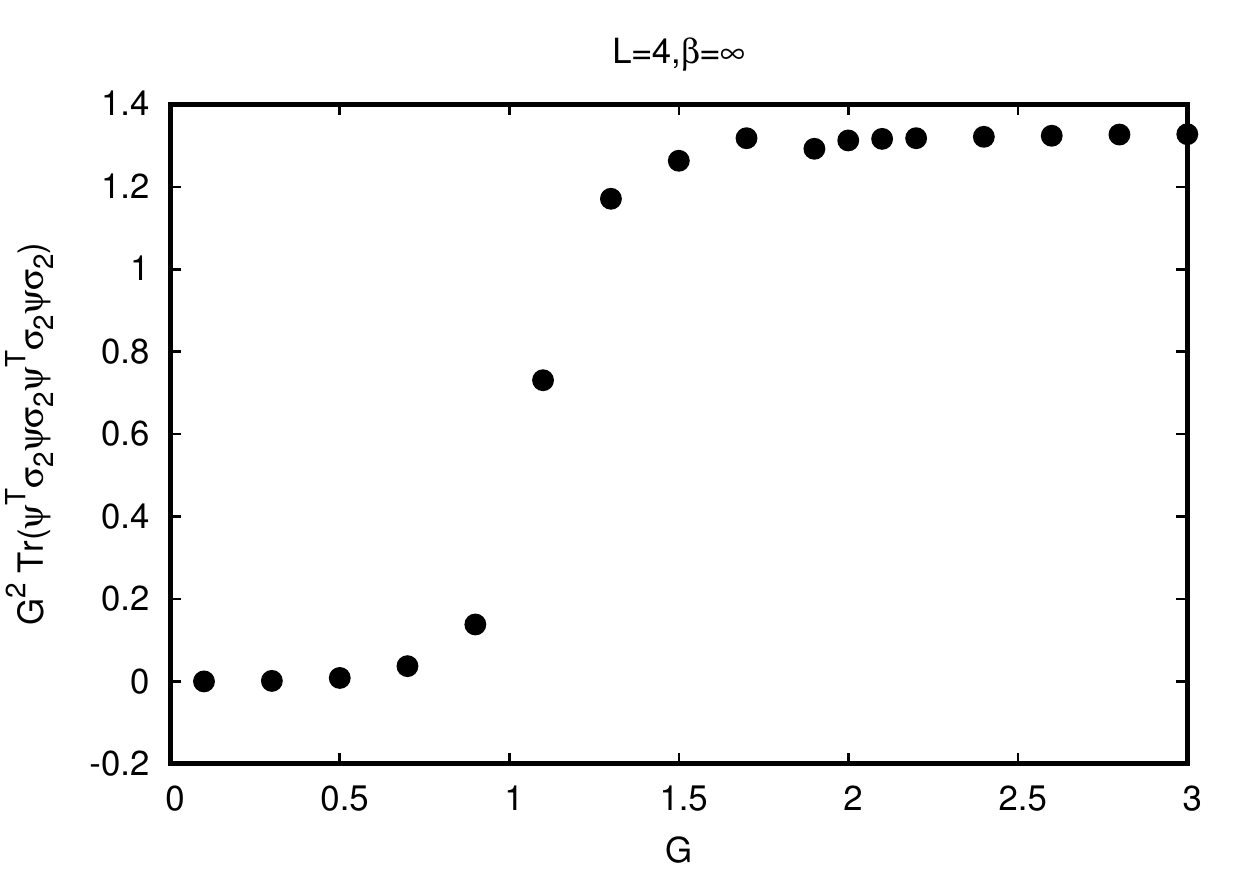} \hfill \includegraphics[width=.5\textwidth]{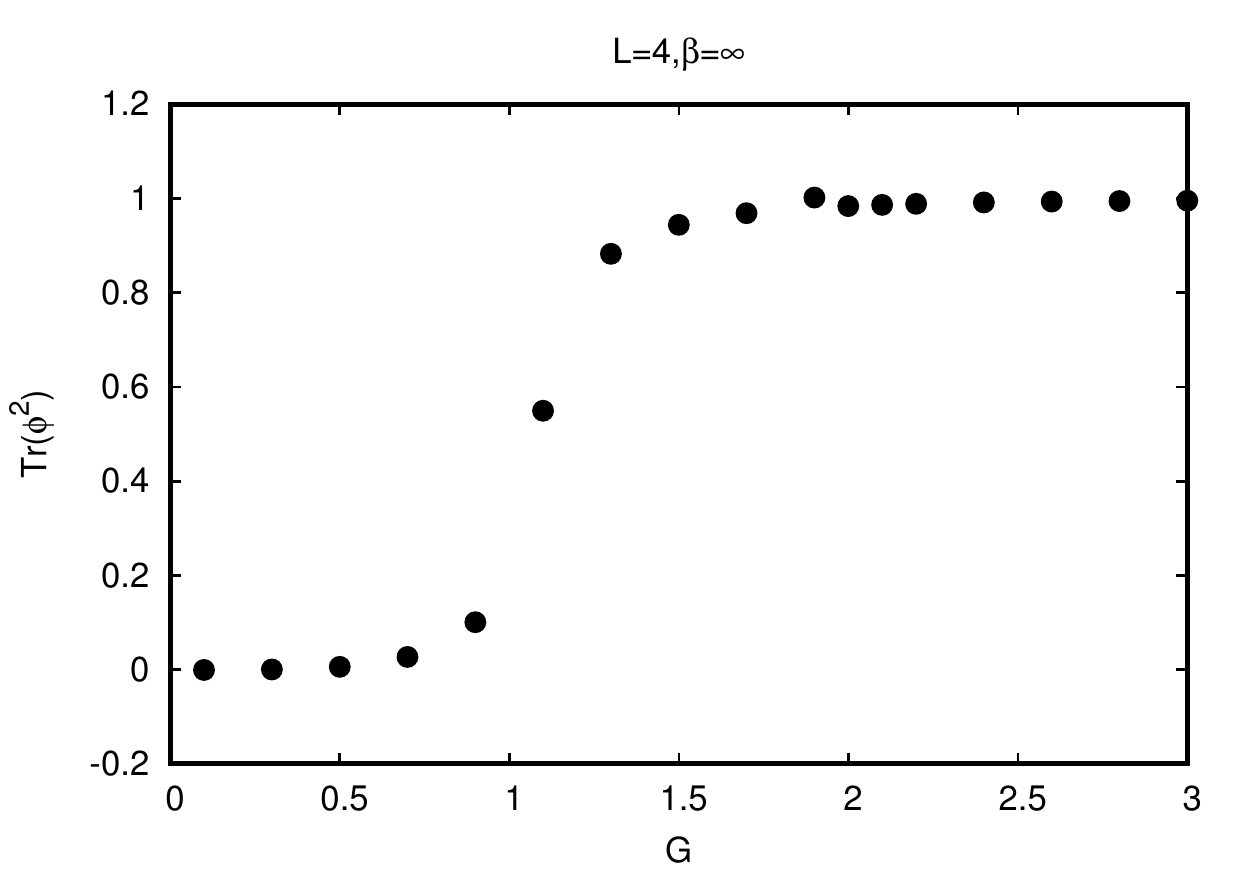}
\caption{ Four-fermion condensate(left) and $ Tr(\phi^2) $ (right) vs $G$ with $ \beta= \infty$ for $ L=4$ }
\label{fig:four}
  \end{figure*}

\begin{figure*}[tbp]
\centering
\includegraphics[width=.5\textwidth]{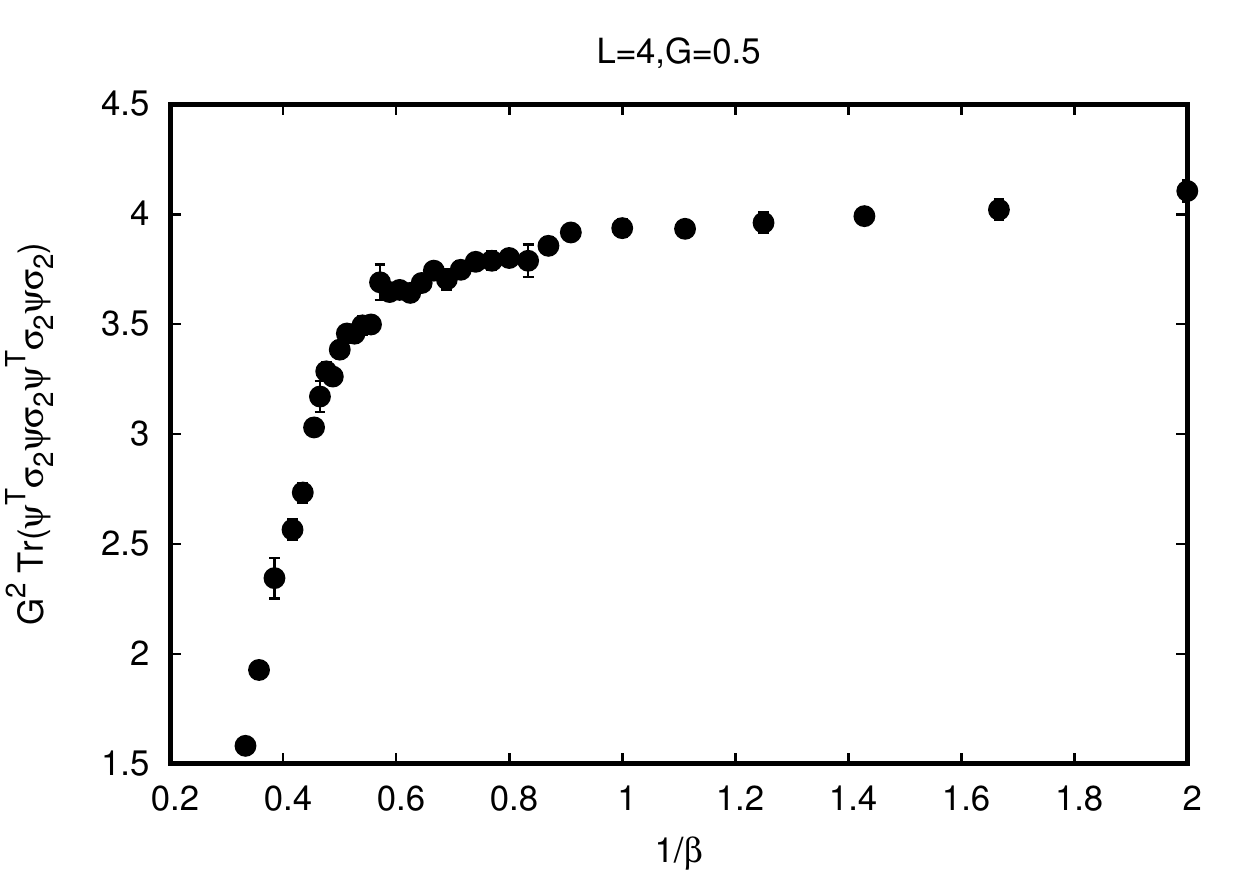}\hfill \includegraphics[width=.5\textwidth]{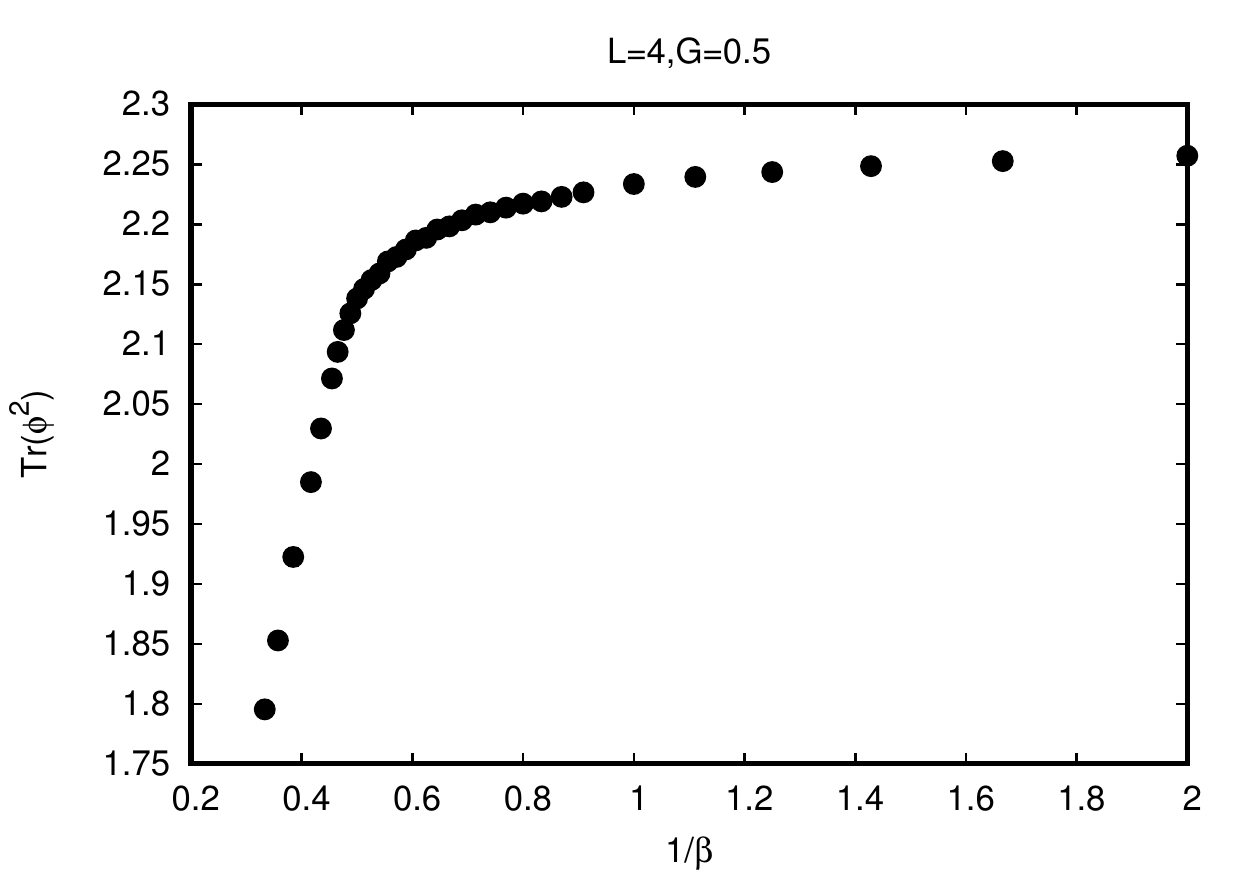}
\caption{ Four-fermion condensate(left) and $ Tr(\phi^2) $ (right) vs $1/\beta$ with $ G=0.5 $ for $ L=4$ }
\label{fig:four_g}
  \end{figure*}

One important test of our code is whether we recover the known behavior of
the $ SO(4)$ model in the weak gauge coupling limit. For this purpose we switch off the gauge fields and scan the four fermion condensate as a function of $G$.This is plotted in Fig.~\ref{fig:four}(left). Another proxy observable for the four fermion (massive) phase is $ Tr (\phi^2) $ which is plotted in Fig.~\ref{fig:four}(right). This plots are consistent with those
reported in ~\cite{Catterall:2016dzf}.
Of course our main goal is to try to realize a four fermion phase through gauge interactions. As a first step in this direction we plot in Fig.~\ref{fig:four_g} the four fermion condensate and $ \langle Tr (\phi^2)  \rangle $ vs $ 1/\beta $ at $ G=0.5 $. At strong gauge coupling the four fermion condensate rises rapidly to a non-zero constant value. Notice that this value of $G$ would not be sufficient to realize a four fermion phase in the absence of gauge interactions. It remains to map out this transition line for a range
of $G$ (using much larger lattice volumes).

\section{Summary}

In this paper we have embedded an $ SO(4) $ invariant four-fermion model into a gauge model by gauging one the two $ SU(2) $ subgroups of $SO(4)$. 
We have described in some detail the construction of the model in terms
of a staggered fermion field transforming in the bifundamental representation 
of $SU(2)\times SU(2)$ and how this representation can be related to the original fundamental representation of $SO(4)$ by imposing a suitable
reality condition on the fermions. Our numerical work is
only just beginning but confirms that the model does indeed reduce to the
original four fermion model in the limit the gauge coupling is sent to zero.
Once the gauge coupling is turned on we have so far only run with a single value of the Yukawa coupling on a small lattice but our results are compatible with the appearance of a four
fermion phase driven in part by strong gauge interactions. We interpret the
four fermion phase as the (gauge invariant) signal of a Higgs phase in
the gauge theory.

Future work will aim to map out the phase diagram of the model using much larger lattices and understand the nature of any critical lines encountered.
It will be particularly interesting to see whether the four fermion phase
survives the limit in which the Yukawa coupling is sent to zero.

\end{document}